\documentstyle{article}

\def\be{\begin{eqnarray}}
\def\ee{\end{eqnarray}}
\def\nn{\nonumber}

\hoffset= - 1.0in         

\begin{document}

\hfill ITEP-TH-28/05

\bigskip

\centerline{\Large{Comment on the Surface Exponential for Tensor Fields
}}

\bigskip

\centerline{\it E.T.Akhmedov, V.Dolotin and A.Morozov}

\bigskip

\centerline{ITEP \& JINR,  Russia}

\bigskip

\centerline{ABSTRACT}

\bigskip

Starting from essentially commutative exponential map $E(B|I)$ for
generic tensor-valued $2$-forms $B$, introduced in \cite{Akh}
as direct generalization of the ordinary non-commutative $P$-exponent
for 1-forms with values in matrices (i.e. in tensors of rank 2),
we suggest a non-trivial but multi-parametric exponential
${\cal E}(B|I|t_\gamma)$, which can serve as an interesting
multi-directional evolution operator in the case of higher ranks.
To emphasize the most important aspects of the story, construction is
restricted to backgrounds $I_{ijk}$, associated with the structure
constants of {\it commutative} associative algebras, what makes it
unsensitive to topology of the $2d$ surface.
Boundary effects are also eliminated (straightfoward generalization
is needed to incorporate them).

\bigskip

\bigskip

\section{Motivations}


Given a matrix-valued one-form $A_{ij}^\mu(x)dx_\mu$ on a line, one can
introduce an ordered exponent, $P\exp \left(\int A_{ij}(x)dx\right)$,
which can be also defined as a limit
\be
z(\hat I+\hat A) \equiv P\exp\left(\int A_{ij}(x)dx\right) =
\lim_{N\rightarrow \infty}
\prod_{n=1}^N\left( \hat I + \frac{1}{N}\hat A\left({n}/{N}\right)\right)
\label{Pexp}
\ee
with the unit matrix $I_{ij} = \delta_{ij}$ (hats denote tensors with
suppressed indices).

When a $1d$ line (real curve) is substituted by $2d$ surface (complex curve)
$\Sigma$, appropriate generalization of $P$-exponent is badly needed for
numerous string theory considerations.
The problem is known under many names, from topological models \cite{TT,VF}
to Connes-Kreimer theory \cite{CK}-\cite{Mal}
and that of the $2$-categories \cite{2cat}.
The role of a chain in (\ref{Pexp}) is now played by a "net" -- a {\it dual}
graph $\Gamma$, describing simplicial complex, which "triangulates"\footnote{
In principle, to imitate surface integrals with arbitrary measures
one needs triangulations with lengths, ascribed to the links \cite{Akhm}.
However, as usual in the matrix-model realizations of string theory
\cite{Mig}, one can ignore the lengths
(consider "equilateral triangulations") -- like
it is done in (\ref{Pexp}) -- and associated quantities will be quite
informative (perhaps, even {\it exhaustively} informative: for example,
it is believed -- and confirmed by numerous calculations of particular
quantities -- that Polyakov's (sum over $2d$ metrics) and Migdal's
(matrix-model sum over equilateral triangulations)
descriptions of string correlators are equivalent). Still, the
interrelation between arbitrary and equilateral triangulations remains an
open problem, touching the fundamental questions of number theory \cite{LM}.
The problem is also known as that of continuous limits in matrix models.
}
$\Sigma$, and tensors $\hat T^{(m)} $ of rank $m$ stand as coupling constants
at the vertices of valence $m$ of this graph.
To provide triangulation of a surface, the graph
should have many vertices with valencies $m\geq 3$,
and essential generalization of (\ref{Pexp}) is unavoidable.
Indices of $\hat T$'s at the vertices are contracted by special
rank $2$ tensors $\hat g$, called metric or propagator (it is assumed to
have upper indices if all $\hat T$ have lower ones).
With such a graph one associates a partition function
\cite{GMS} $\hat Z_\Gamma^{\hat g}\{\hat T\}$ which is a tensor of rank
${\rm ext}(\Gamma)$, equal to the number of external legs of the graph
$\Gamma$ (if $\Sigma$ has no boundaries, ${\rm ext}(\Gamma)=0$).

\section{Suggestion of ref.\cite{Akh}
}

\noindent

In \cite{Akh} a principally important step is made to bring the abstract
constructions of \cite{GMS} and \cite{D}-\cite{Ja} closer to appropriate
generalization of (\ref{Pexp}). The crucial additional structure,
used in (\ref{Pexp}), is decomposition of the rank-2 tensor
$\hat t = \hat I + \hat A$ into "background" $\hat I$ and "dynamical"
$\hat A$ parts. $z(\hat I) = 1$ is trivial, though $I_{ij} =
\delta_{ij} \neq 0$ itself is not.

For complex curves the background $\hat I$ should necessarily have
rank greater than $2$, for graphs of valence $3$ it should have rank $3$.
It is well known \cite{VF}, that there are non-trivial $\hat I$ and
associated $\hat g$ exist with trivial $$Z_\Gamma^{\hat g}(\hat I) = 1$$
for {\it all} $\Gamma$ without external legs\footnote{
Of course,  for ${\rm ext}(\Gamma)\neq 0$ the
$\hat Z_\Gamma^{\hat g}(\hat I)$
is an operator (has external indices) and can not be unity. However, it
can be made dependent only on ${\rm ext}(\Gamma)$ but not on $\Gamma$
itself, see s.\ref{Al} below.
}
(they can be build
from structure constants of any commutative associative algebra, see
s.\ref{Al} below, but this does not exhaust all the possibilities),
and ref.\cite{Akh} suggests to make use of them exactly in the same way as
in (\ref{Pexp}). Since in this construction $\hat g$ is rigidly
linked to  $\hat I$, we suppress the $\hat g$ labels in most formulas below.

According to the definition of $\hat Z_\Gamma$ for
$\hat T = \hat I + \hat B$
we have:\footnote{Note that the graph automatically picks up
the tensors of appropriate rank from $\hat B$, if there is no match,
$\hat Z = 0$. If we assume that $\hat I$ is exactly of rank $3$ while
$\hat B$ consists of tensors of various ranks (we"ll see below that it
is useful not to restrict $\hat B$ to rank $3$ only), then only
subgraphs $\Gamma/\gamma$ with vertices of valence $3$ will
contribute to the sum. The subgraph $\Gamma/\gamma$ is defined
by throwing away all the vertices
of $\gamma$ and all the links between them, ${\rm ext}(\Gamma/\gamma) =
{\rm ext}(\Gamma) + {\rm ext}(\gamma) - 2(\#\ {\rm of\ common\ external\
legs\ of\ }\Gamma\ {\rm and}\ \gamma)$). In other words, $\gamma$ is
treated as "vertex-subgraph" of $\Gamma$. As explained in \cite{GMS},
the "vertex-subgraphs" (in variance with the "box-subgraphs") are related
to relatively simple set-theoretical aspect of quantum field theory
(to the $Shift\ {\cal M}$ rather than $Diff\ {\cal M}$ structure of diagram
technique). In the present context this is the reason behind the
over-simplicity (commutativity property (\ref{exppro})) of the exponential
(\ref{limB}).
}
\be
\hat Z_\Gamma(\hat I + \hat B) = \sum_{\gamma \subset \Gamma}
\hat Z_{\Gamma/\gamma}(\hat I) \hat Z_\gamma(\hat B)
\ee

If we now take a limit of large $|\Gamma| \equiv (\# \ {\rm of \ vertices\ in}\
\Gamma)$, with graph growing in both dimensions to form a dense net,
and look at the terms with a given power of $B$, then
statistically only graphs $\gamma$ consisting of isolated points will
survive after appropriate rescaling of $B$, and this logic leads to the
following generalization of (\ref{Pexp}) \cite{Akh}:
\be
\hat E(\hat B|\hat I) = \lim_{|\Gamma|\rightarrow\infty}
\hat Z_\Gamma\left(\hat I + \frac{1}{|\Gamma|}\hat B\right)
\label{limB}
\ee
(the argument $\hat I$ in $\hat E$ will often be suppressed below).
This is a very nice and interesting quantity, but it is essentially
Abelian: as we shall see in (\ref{exppro}),
$\hat E(\hat B_1)\hat E(\hat B_2) =\hat E(\hat
B_1+\hat B_2)$ (for example, for a rank-$3$ $2$-form
$\hat B = B_{ijk}^{\mu\nu} dx_\mu \wedge dx_\nu$ on $\Sigma$
we can define a surface integral as
$E\left(\int_\Sigma \hat B\right)$ and never encounter any ordering
problems). This happens for the same reason that the homotopic groups
$\pi_k$ are commutative for $k>1$: any two insertions of $B$
at two {\it remote} points can be easily permuted, moving one {\it around}
another.

In what follows we give a more formal description of above construction,
getting rid of a subtle limiting procedure in (\ref{limB}),
and introduce -- with the help of $\hat E(\hat B)$, just changing its
argument -- a less trivial exponential $\hat {\cal E}(\hat B|t)$.
It should be useful in applications, it is well defined,
but no "conceptual" limiting formula like (\ref{limB})
is immediately available for it.

\section{Backgrounds $\hat I$ from commutative associative algebras
${\cal A}$ \label{Al}}

\noindent

Let $\left(\check C_i\right)^k_j = C^k_{ij}$ be
structure constants of an associative algebra ${\cal A}$
($\phi_i * \phi_j = C_{ij}^k\phi_k$):
\be
\left[\check C_i, \check C_j\right] = 0
\label{assorel}
\ee
Introduce a set of {\it symmetric} tensors $\hat I^{(n)}$:
\be
I_{i_1\ldots i_n} \equiv {\rm Tr}
\Big(\check C_{i_1}\ldots \check C_{i_n}\Big)
\label{Iten}
\ee
Among them will be the {\bf metric}\footnote{
Note that this choice of metric is {\it different} from 
$G_{ij}^{(m)} = I^{(3)}_{ijm}$,
used in the context of the generalized WDVV equations in \cite{WDVVMMM}.
}
\be
g_{ij} = I_{ij} =
{\rm Tr} (\check C_{i}\check C_{j})
\label{met}
\ee
and the {\bf elementary vertex} $I_{ijk} = {\rm Tr}
(\check C_{i}\check C_{j}\check C_{k})$.
Metric will be assumed non-degenerate and its inverse $g^{ij}$
will be used to raise indices.

In what follows we impose additional {\bf commutativity}
condition on the structure constants:
\be
C^k_{ij} = C^k_{ji}
\label{coco}
\ee
Then tensors $\hat I$ are not just cyclic, but totally symmetric.

{\bf Lemma:} For commutative associative algebra ${\cal A}$
$$I_{ijk} = g_{im}C^m_{jk}$$
or simply $I^m_{jk} = C^m_{jk}$. Indeed,
$$g_{im}C^m_{jk} = C_{il}^nC_{mn}^lC_{jk}^m
\ \stackrel{(\ref{coco})}{=}\
C_{il}^nC_{nm}^lC_{jk}^m \ \stackrel{(\ref{assorel})}{=}\
C_{il}^nC_{jm}^lC_{nk}^m = I_{ikj} \ \stackrel{(\ref{coco})}{=}\
I_{ijk}
$$

\bigskip

{\bf Lemma:} For commutative associative algebra ${\cal A}$
\be
I_{i_1\ldots i_m k_1\ldots k_r}I_{j_1\ldots j_n}^{k_1\ldots k_r} =
I_{i_1\ldots i_m k_1\ldots k_r}
I_{j_1\ldots j_n \tilde k_1\ldots \tilde k_r}
g^{k_1\tilde k_1}\ldots g^{k_r\tilde k_r} =
I_{i_1\ldots i_m j_1\ldots j_n}
\ee
for any $r\neq 0$ and any $m$ and $n$.
The proof follows from the observation that (\ref{assorel})
and (\ref{met}) provide the two transformations
(the "flip" or "zigzag" transform and tadpole-eliminator),
which generate a group with transitive action on the space of all
connected triangulations (see \cite{Akh} for the relevant illustrations).

\bigskip

Obviously, for any {\it connected} $\Gamma$,
$\hat Z_\Gamma(\hat I)$ is given by these tensors $I$:
\be
\left(Z_\Gamma^{\hat g}(\hat I)\right)_{i_1\ldots i_{{\rm ext}(\Gamma)}}
= I_{i_1\ldots i_{{\rm ext}(\Gamma)}}
\label{Zind}
\ee
In other words, in background theory the connected
diagram depends only on the number of external legs.
One can of course associate additional factors with graphs,
counting the numbers of vertices and loops, but they do not depend on
${\cal A}$ and are not of immediate interest for our consideration.

If $\Gamma$ consists of disconnected parts, $\hat Z_\Gamma(\hat I)$ will
be a tensor product of $\hat I$-tensors.

If other {\it traceless} tensors $B^{(m)}$ of rank $m$ are allowed in the
vertices, we get a non-trivial $B$-theory in the ${\cal A}$-background.
Original $\hat Z(\hat T)$ provides the unified background-independent
formulation. Still, explicit transformation from one background to another
remains an interesting open problem.

\section{Commutative exponential
}

\noindent

Introduce a $\hat B$-dependent tensor of rank $n$
\be
E^{(n)}_{i_1\ldots i_n}(\hat B) =
\sum_{m=1}^\infty \sum_{\{r_m\}}\sigma\{r_m\}
\left\{
\left(B^{j^{(1)}_1}\ldots B^{j^{(1)}_{r_1}}\right)
\left(B^{j^{(2)}_1j^{(2)}_2}\ldots
B^{j^{(2)}_{2r_2-1}j^{(2)}_{2r_2}}\right)
\ldots \right.\nn \\ \left. \ldots
\left(B^{j^{(m)}_1\ldots j^{(m)}_m}\ldots
B^{j^{(m)}_{r_m-m+1}\ldots j^{(m)}_{r_m}}\right)
\ldots\right\}
I_{i_1\ldots i_k j^{(1)}_1\ldots j^{(m)}_{r_m}\ldots}
\label{Abexp}
\ee
Here $\hat B = \{\hat B^{(1)}, \hat B^{(2)},\ldots, \hat B^{(m)},
\ldots\}$ is a direct sum of tensors of all possible ranks and
the sum in (\ref{Abexp}) is over all possible sets,
including any number $r_m$ of tensors of rank $m$.

{\bf Theorem:}
For appropriate choice of the combinatorial factors $\sigma\{r_m\}$
the map $\hat E(\hat B)$ satisfies the exponential property:\footnote{
Note that in terms of $\hat T = \hat I + \hat B$ this relation does
not look homogeneous:
$$\hat Z_\Gamma(\hat I + \hat B_1 + \hat B_2) =
\hat Z_\Gamma(\hat I+\hat B_1)\hat Z_\Gamma(\hat I+\hat B_2) +
O\left(|\Gamma|^{-1}\right),$$
and holds because of the $\Gamma$-independence property (\ref{Zind})
of $\hat Z_\Gamma(\hat I)$.
}
\be
E^{(m+r)}_{i_1\ldots i_m k_1\ldots k_r}(B_1)
{E^{(n+r)}}_{j_1\ldots j_n}^{k_1\ldots k_r}(B_2) =
E^{(m+n)}_{i_1\ldots i_m j_1\ldots j_n}(B_1+B_2)
\label{exppro}
\ee
for any $r\neq 0$ and any $m$ and $n$ (no sum over $r$ is taken).
The relevant choice of $\sigma\{r_m\}$ is the usual Feynman-diagram
factorial (see, for example, the "generalized Wick theorem" in
\cite{amm23})
\be
\sigma\{r_m\} = \prod_{m=1} \frac{1}{r_m!(m!)^{r_m}}
\ee
The factors $m!$ can be eliminated by rescaling of $B^{(m)}$.
As immediate corollary of (\ref{exppro}), derivative of the exponent
functor $\hat E(\hat B)$ is $\hat E(\hat B)$ itself:
\be
\delta E^{(n)}_{i_1\ldots i_n}(B) =
\sum_{m=1}^\infty E^{(n+m)}_{i_1\ldots i_n j_1\ldots j_m}(B)
\delta B_{(m)}^{j_1\ldots j_m} + O(\delta B^2)
\ee

Thus nothing like non-trivial Campbell-Hausdorff formula
\cite{CH} (which describes the product of $P$-exponents (\ref{Pexp}))
arises for $\hat E(\hat B)$, potential non-commutativity of tensor
product is completely eliminated by the naive continuum limit
(\ref{limB}), as a corollary of the relation (\ref{Zind}) and the
possibility to rely upon {\it connected} graphs.

\section{Non-trivial exponential and directions of immediate generalization}

\noindent

Potential origin of non-triviality of the exponential
is two-fold: there can be contributions from non-trivial
(not totally disconnected) subgraphs $\gamma$ to $\hat Z_\gamma(\hat B)$
and from {\it disconnected} factor-graphs $\Gamma/\gamma$ to
$\hat Z_{\Gamma/\gamma}(\hat I)$.
The first origin (contribution from non-trivial $\gamma$) is eliminated
by naive continuum limit -- both in $1d$ formula (\ref{Pexp}) and in the
$2d$ one (\ref{Abexp}). In $1d$ the property (\ref{Zind}) perfectly holds
for {\it connected} graphs, but disconnected $\Gamma/\gamma$ also contribute
to (\ref{Pexp}). What happens in $2d$ is that disconnected $\Gamma/\gamma$
are statistically damped in the naive continuum limit, {\it together}
with non-trivial $\gamma$ -- and direct generalization of the
non-Abelian (\ref{Pexp}) from lines to surfaces is Abelian (commutative)!

In order to get a non-commutative exponential in $2d$ one can, however,
revive the contributions from {\it non-trivial} $\gamma$, simply by
introducing a non-trivial multi-time evolution operator:
\be
\hat {\cal E}(B|t) = \hat E\left(\sum_{{\rm connected}\ \gamma} t_\gamma
\hat Z_\gamma(B)\right)
\label{calE}
\ee
(of course, one can do -- and often does -- the same in $1d$).
Similarly, one can add contributions from disconnected $\Gamma/\gamma$
by introducing certain non-local operators (involving contour integrals)
in the exponent. Despite such quantities may seem less natural than
(\ref{limB}), they naturally arise in physically relevant evolution
operators and even in actions, bare and effective. Moreover, for
special $\hat B$, for example, totally antisymmetric, the leading
contribution with single-point $\gamma$ vanishes in {\it symmetric}
background $\hat I$. Then the next-to-leading
contribution -- from single-link (and two-point) $\gamma$ -- can be
described by appropriately modified limiting prescription (\ref{limB}).

$\hat {\cal E}(\hat B|\hat I|t)$ is already a non-trivial (operator)
special function, which deserves attention and investigation.
Note, that even if $\hat B$ was a rank-$3$ tensor,
and all the relevant graphs were of valence $3$, the
tensors $\hat Z_\gamma(\hat B)$ which contribute to the argument of
$\hat E$ in (\ref{calE}) have ranks ${\rm ext}(\gamma)$, not obligatory
equal to $3$.

Despite non-trivial graphs $\gamma$ are now incorporated in (\ref{calE}),
they are still restricted to {\it lie inside} $\Sigma$, remain separated
from the
boundary of $\Sigma$ by requirement of connectedness of $\Gamma/\gamma$
(external legs of $\Gamma$ are not allowed to belong to $\gamma$).
Additional corrections to (\ref{calE}) are needed to make it sensitive
to boundary effects.

In this note we restricted consideration to the simplest possible case
of commutative algebra ${\cal A}$, when tensors $\hat I$ are totally
symmetric. Relaxing this requirement, one gets $\hat I$ with only cyclic
symmetry, then (\ref{Zind}) gets more complicated: universality classes
are no longer enumerated by ${\rm ext}(\Gamma)$, dependence on the number
of handles arises and description in terms of fat graphs is needed
\cite{Akh} (see \cite{Mal} for the corresponding generalization of
\cite{GMS}).

An interesting part of this story is exponentiation
(the algebra $\rightarrow$ group lifting)
of associative algebras and higher-rank multiplications
\cite{CD}. It involves limits like (\ref{limB}) along peculiar chains
of graphs (obtained, for example, by iterative blowing up of triple
vertices into triangles). In such situations, the $B$ insertions have
higher probability to break the graph into disconnected components than
for generic net-graphs. Still, enhancement is not sufficient and,
like in (\ref{limB}), such contributions
remain statistically damped in naive continuum limit. Therefore
transition from (\ref{limB}) to (\ref{calE}) should still
be made "by hands".

Related open question concerns generalization of background $\hat I$
from the rank-$3$ case (related to associative algebras) to generic
situation and connection of this problem to Batalin-\-Vilkovisky theory of
Massis operations \cite{BV}-\cite{GL}.

\section{Acknowledgements}

\noindent

We acknowledge discussions with A.Gerasimov, A.Losev and A.Rosly.

Our work is supported by Federal Program of the Russian Ministry
for Industry, Science and Technology No 40.052.1.1.1112 and by
RFBR grants 04-02-16880 (E.A. \& A.M.) and
04-02-17227 (V.D.).

\end{document}